\documentclass[12pt]{article}

\usepackage{scicite}

\usepackage{times}

\usepackage{caption}

\usepackage{graphicx}

\usepackage{amsmath}

\newenvironment{sciabstract}{%
\begin{quote} \bf}
{\end{quote}}
\title{Generation of entanglement using a short-wavelength seeded free-electron laser} 
\author{
\parbox{\linewidth}{\centering
Saikat Nandi$^{1,\ast}$, Axel Stenquist$^{2,\dagger}$, Asimina Papoulia$^{2,\dagger}$, Edvin Olofsson$^2$, Laura Badano$^{3}$, Mattias Bertolino$^2$, David Busto$^{2}$, Carlo Callegari$^{3}$, Stefanos Carlstr\"{o}m$^{2}$, Miltcho B. Danailov$^{3}$, Philipp V. Demekhin$^{4}$, Michele Di Fraia$^{3}$, Per Eng-Johnsson$^{2}$, Raimund Feifel$^{5}$, Guillaume Gallician$^{6}$, Luca Giannessi$^{3,7}$, Mathieu Gisselbrecht$^{2}$, Michele Manfredda$^{3}$, Michael Meyer$^{8}$, Catalin Miron$^{6,9}$, Jasper Peschel$^{2}$, Oksana Plekan$^{3}$, Kevin C. Prince$^{3}$, Richard J. Squibb$^{5}$, Marco Zangrando$^{3,10}$, Felipe Zapata$^{2}$, Shiyang Zhong$^{2}$, Jan Marcus Dahlstr\"{o}m$^{2,\ast}$}
\\
\\
\normalsize{$^{1}$Universit\'{e} de Lyon, Universit\'{e} Claude Bernard Lyon 1, CNRS,}\\
\normalsize{Institut Lumi\`{e}re Mati\`{e}re, 69622, Villeurbanne, France}\\
\normalsize{$^{2}$Department of Physics, Lund University, 22100 Lund, Sweden}\\
\normalsize{$^{3}$Elettra-Sincrotrone Trieste, 34149 Basovizza, Trieste, Italy}\\
\normalsize{$^{4}$Institute of Physics and CINSaT, University of Kassel, 34132 Kassel, Germany}\\ 
\normalsize{$^{5}$Department of Physics, University of Gothenburg, 41258 Gothenburg, Sweden}\\
\normalsize{$^{6}$Universit\'{e} Paris-Saclay, CEA, CNRS, LIDYL, 91191 Gif-sur-Yvette, France}\\
\normalsize{$^{7}$INFN, Laboratori Nazionali di Frascati, 00044 Frascati, Italy}\\
\normalsize{$^{8}$European XFEL, 22869 Schenefeld, Germany}\\
\normalsize{$^{9}$ELI-NP, ``Horia Hulubei'' National Institute for Physics and Nuclear Engineering,}\\
\normalsize{077125 M\v{a}gurele, Romania}\\
\normalsize{$^{10}$IOM-CNR, Istituto Officina dei Materiali, 34149 Basovizza, Trieste, Italy}\\
\\
\normalsize{$^\dagger$These authors contributed equally.}\\
\normalsize{$^\ast$Corresponding author. E-mail:  saikat.nandi@univ-lyon1.fr; marcus.dahlstrom@matfys.lth.se}
}

\date{}

\begin{document}

\baselineskip24pt

\maketitle 

\begin{sciabstract} 

Quantum entanglement between the degrees of freedom encountered in the classical world is challenging to observe due to the surrounding environment. To elucidate this issue, we investigate the entanglement generated over ultrafast timescales in a bipartite quantum system comprising two massive particles: a free-moving photoelectron, which expands to a mesoscopic length-scale, and a light-dressed atomic ion, which represents a hybrid state of light and matter. Although the photoelectron spectra are measured classically, the entanglement allows us to reveal information about the dressed-state dynamics of the ion and the femtosecond extreme ultraviolet pulses delivered by a seeded free-electron laser. The observed generation of entanglement is interpreted using the time-dependent von Neumann entropy. Our results unveil the potential for using short-wavelength coherent light pulses from free-electron lasers to generate entangled photoelectron and ion systems for studying spooky action at a distance.

\end{sciabstract}

\section*{Introduction}

When two or more quantum particles in a many-body system are entangled, the system's wavefunction cannot be factorized as a product of the wavefunctions of its constituents\cite{horodecki2009}. Entanglement has now been established as the driving force behind the stunning development of quantum information science\cite{mikeike2010,yin2017,arute2019,zhong2020,hermans2022,krinner2022}. A key example of entanglement emerges in Einstein's photoelectric effect, where photoelectrons and ions form a bipartite system as measurements of the electron's kinetic energy provides information about the state of the ion. It offers an excellent opportunity to probe entanglement between massive particles, namely the photoelectron and photoion. 
Even though entanglement in photoinization was initially studied in systems interacting with soft x-rays from synchrotron radiation sources\cite{akoury2007,schoffler2008}, it has now become a subject of intensive research in the time domain\cite{smirnova2010,goulielmakis2010} and recently, attosecond control of entanglement was achieved\cite{vrakking2021,koll2022}. 
The emergence of decoherence in attosecond interferometric experiments has been investigated for atomic targets\cite{bouchet2020,busto2022}. Of particular interests are the quantum electrodynamical aspects of strong-field laser-matter interaction, such as the high-order harmonics generation process\cite{stammer2023,bhattacharya2023,tzur2023,gorlach2023}. The production of non-classical states\cite{tsatrafyllis2017,tsatrafyllis2019}, photonic entanglement\cite{stammer2022}, photon-atom entanglement\cite{gorlach2020}, and macroscopic superposition via creation of Schr\"{o}dinger cat states\cite{lewenstein2021} in high-order harmonic generation have received considerable attention lately. Additional examples include the investigations of entanglement between different degrees of freedom in atomic photoionization\cite{majorosi2017,maxwell2022} and preparation of entangled atom-pairs in molecular dissociation\cite{eckart2023}. In this context, the emergence of seeded free-electron lasers (FELs), such as FERMI\cite{allaria2012}, provides a unique niche to investigate coherent light-matter interaction using intense femtosecond pulses in the extreme ultraviolet (XUV) spectral range. Such XUV pulses from FERMI, with high spectral and temporal coherence, have already been employed to drive Rabi oscillations\cite{nandi2022} and to attain attosecond coherent control\cite{prince2016,maroju2023}. 
Our work addresses the question whether these intense short-wavelength light pulses could be used to generate entanglement between ions and electrons in photoinization processes.

According to Einstein's equation of the photoelectric effect, the kinetic energy of photoelectrons is given by: $E^\mathrm{kin}=\hbar\omega-E^\mathrm{bin}$, where $\hbar$ is the reduced Planck constant and $\omega$ is the angular frequency of the electromagnetic radiation interacting with a system that has binding energy $E^\mathrm{bin}$. In multi-electronic systems, photoemission may result in different ion channels: $i$ or $j$. 
If the ion channels correspond to different binding energies: $E^\mathrm{bin}_i \ne E^\mathrm{bin}_j$, the energy conservation of the ionization processes makes the associated photoelectrons distinguishable by their kinetic energies\cite{pabst2011,nishi2019}: $E^\mathrm{kin}_{i}\ne E^\mathrm{kin}_{j}$. The final state (${\cal F}$) is then entangled because the wavefunctions of the ion (${\cal I}$) and the photoelectron (${\cal P}$) can not be factorized\cite{ruberti2022}, i.e., $|\Psi^{({\cal F})}\rangle \ne |\psi^{({\cal I})}\rangle \otimes |\psi^{(P)}\rangle$.
This result is a particular case of the general superposition of product states:  
\begin{equation}
|\Psi^{({\cal F})}\rangle\,=\,\sum_i  \int_0^\infty dE^\mathrm{kin} \,\, 
\alpha_{i}(E^\mathrm{kin})
|\psi_i^{({\cal I})}\rangle\otimes|\psi^{({\cal P})}(E^\mathrm{kin})\rangle,
\label{eq:eq1}
\end{equation}
where $\alpha_{i}(E^\mathrm{kin})$ are the complex amplitudes of the final state. In this way, a bipartite system of the combined Hilbert spaces of the ion, and the photoelectron, builds up the total Hilbert space of the final state\cite{haroche2006}.

Here, we study how a non-entangled (factorizable) state of ion and photoelectron can be driven into a fully entangled (non-factorizable) final state by subsequent dressing of the ion by an intense field. A neutral helium atom, initially in its ground state, $|g\rangle=|1\mathrm{s}^2\rangle$, is ionized by the FEL pulse. The photon energy of the pulse was varied in a range centered around $\hbar\omega= 40.8$ eV, which is high enough for single-photon ionization of the neutral atom, but considerably lower than the single-photon double-ionization threshold\cite{lablanquie1990}. This results in a single ionic channel, $|a\rangle=|1\mathrm{s}^+\rangle$, which is factorizable with a single term in the channel sum: $i=a$ in Eq.~(1). 
However, the photon energy of the FEL is also close to the lowest ionic transition: $\hbar\omega_{ba}=40.814$~eV\cite{kramida2022}, where $|b\rangle=|2\mathrm{p}^+\rangle$ is the lowest dipole allowed excited state of He$^{+}$. As a result, exchange of a second photon from the same FEL pulse can lead to Rabi oscillations in the residual He$^+$ ion. The ensuing Rabi dynamics in the ion have been predicted by Grobe and Eberly to modify the kinetic energy distribution of the emitted photoelectron with the appearance of a doublet structure\cite{grobe1993}. While the phenomenon has been reported for calcium atoms by Walker and co-workers\cite{walker1995}, additional atomic levels resonant with the ultraviolet laser field made the interpretation less clear as compared to the original proposal. In more recent work, Zhang and Rohringer have shown that the effect should be clearly visible for helium atoms, when driven by intense XUV pulses\cite{zhang2014}. The two-electron helium case has been further studied by Yu and Madsen, within a one-dimensional model, where electron-electron correlation was found to be negligible\cite{yu2018}. While the underlying mechanism has been referred to as coherence transfer by electron--electron correlation\cite{grobe1993}, continuum--continuum Autler-Townes splitting\cite{walker1995} and core-resonant ionization\cite{yu2018}, a clear physical picture describing how the phenomenon takes place over time has not been presented. We demonstrate that, in fact, the formation of the Grobe--Eberly doublet is a manifestation of quantum entanglement being generated between the photoelectron and the dressed ion by high intensities of the FEL pulses driving the photoionization processes (see Fig.~1A for a pictorial depiction). Furthermore, the theoretical calculations presented in the current work illustrate that electron-electron Coulomb interaction is not required for the ultrafast formation of the doublet (see Supplementary Materials (abbreviated as SM in the following) for details).

\section*{Results}

Our experiment was performed with XUV pulses from FERMI having an estimated full-width-at-half-maximum (FWHM) $\Delta t^\mathrm{FEL}=73$\,fs (see SM for details about the FEL-pulse properties). The measurements were carried out at three different effective intensities (adjusted with the use of metallic filters) of the FEL pulse, marked as yellow circles in Fig.~1B. To interpret the results, we employed an analytical model that includes photoionization of the He atom with inverse ionization rate, $\tau_g$, and Rabi oscillations of the electronic population in the He$^+$ ion with frequency $\Omega$. The analytical model was further validated by numerical simulations, where the evolution of the neutral He atom and He$^+$ have been coupled using propagation of the time-dependent Dirac equation. This allows us to include strong-field effects, such as the AC Stark shifts, the depletion of the ion, as well as relativistic shifts and splittings of the ionic levels. The photoionization dynamics of the neutral helium atom are described at the level of the relativistic time-dependent configuration interaction singles\cite{zapata2022}, which is a good approximation for the considered FEL photon energy as it is below the double excitation threshold. The inverse ionization rate of the ion, $\tau_{\mathrm{ion}}$, and the Rabi period, $T_{\mathrm{R}}=2\pi/\Omega$, respectively, provide the upper and lower bounds for the dynamics. Generation of quantum entanglement requires $T_R< \Delta t^\mathrm{FEL}$ and $T_{R} \ll \tau_g$ along with FEL-intensities below $10^{15}$ W/cm$^2$ (see Fig.~1B). The spontaneous emission lifetime, $\tau_{\mathrm{SE}}=0.1$ ns, acts as a natural upper limit to the timescale of the overall coherent dynamics.

In Fig.~2A, we display the experimental photoelectron spectra, recorded at the highest FEL intensity. Instead of a monotonically increasing photo-line as a function of the photon energy, predicted by Einstein's equation of the photoelectric effect, the recorded spectra reveal clear hints of an avoided crossing. The dashed lines correspond to the kinetic energies from the analytical model based on the dressed-state picture: 
$E^\mathrm{kin}_\pm =\hbar\omega- E^\mathrm{bin} +\frac{1}{2}\hbar\Delta\omega \pm \frac{1}{2}\hbar W$, where $\Delta\omega$ is the detuning of the photon energy with respect to $\omega_{\mathrm{ba}}$, $E^\mathrm{bin}=\epsilon_a-\epsilon_g$ with $\epsilon_a$ ($\epsilon_g$) being the energy of the ion (atom), and $W=\sqrt{\Omega^2+\Delta\omega^2}$ is the generalized Rabi frequency with effective intensity $I^\mathrm{FEL}=1.25\times10^{13}$ W/cm$^2$. The photoelectron spectra from the analytical model are shown in Fig.~2C for a FEL pulse duration of $73$ fs and peak intensity of $1.25\times10^{13}$ W/cm$^2$. Because photoelectrons originate from a finite interaction region, the theoretical photoelectron spectra for single FEL-intensity have been recalculated by taking the volume-averaging effect into account (see Fig.~2E). Similar photoelectron spectra, measured at an intermediate FEL intensity, $I^\mathrm{FEL}=3.35\times10^{12}$ W/cm$^2$, are shown in Fig.~2B. The theoretical photoelectron spectra at this $I^\mathrm{FEL}$, for both single intensity and volume-averaged contributions, are displayed in Fig.~2, D and F, respectively. The experimental spectra recorded at the lowest FEL intensity, corresponding to $I^\mathrm{FEL}=6.25\times10^{11}$ W/cm$^2$, are shown in the Fig.~S1 (see SM). As expected, no avoided crossing is observed in the absence of entanglement.

To remove the effect of spectral broadening due to the photon bandwidth and the spectrometer resolution, we applied a deconvolution procedure to the measured photoelectron spectra for retrieving any underlying features (see SM for details about the data analysis). For the highest intensity, $I^\mathrm{FEL}=1.25\times10^{13}$ W/cm$^2$, the deconvoluted photoelectron maps are shown in Fig.~3A. A twisted structure along with weak wing-like features is observed, which are absent in the theoretical photoelectron maps in Fig.~2E. 
We interpret this as a mixture between the entangled and non-entangled photoelectron spectra, which follows from the theoretical photoelectron spectra in Fig.~3B, containing both these contributions (at $I^\mathrm{FEL}=1.25\times10^{13}$ W/cm$^2$), thus displaying a clear twist in it. Guided by this finding, we fitted the deconvoluted spectra at each photon energy with several Voigt functions to extract the two different contributions (see SM). The non-entangled contribution is shown in Fig.~3C, which displays a monotonically increasing photo-line, as predicted by Einstein's equation of the photoelectric effect (dashed line). Fig.~3D shows the entangled contribution, as obtained from the difference between the deconvoluted spectra and the non-entangled part. An avoided crossing as a function of the photon energy emerges as a signature of the dressed ion dynamics in He$^+$. The dashed lines correspond to the kinetic energies from the analytical model with $I^\mathrm{FEL}=1.25\times10^{13}$ W/cm$^2$, same as in Fig.~2A. There is an excellent agreement between the entangled experimental spectra in Fig.~3D and its theoretical counterpart in Fig.~2E. 
The Fig.~3D provides a direct experimental proof of the manifestation of quantum entanglement between the dressed ion and the photoelectron - the information about coherent oscillations in the former can be read from the latter. In the absence of quantum entanglement, however, the emitted photoelectron has no knowledge of the Rabi oscillations in the ion (see Fig.~3C). 

\section*{Discussion}
In our experiment, the XUV pulse having $\Delta t^\mathrm{FEL}= 73$ fs is the entangler. It drives Rabi oscillations within $T_R= 21.6$ fs in the ion, while photoionization with inverse rate of $\tau_g\approx 172$~fs from the atom is still taking place (see Fig.~1B). Since, $T_R< \Delta t^\mathrm{FEL}< \tau_g$, this leads to the generation of entanglement between the  photoelectron and the ion. 
The emitted electron is a free particle, described by a wavepacket spreading rapidly in space as a function of time. To get an idea about the size of the photoelectron wavepacket, we calculated the classical limit arising from the product of the speed of the photoelectron as a point particle and the interaction time between the entangler and the residual ion. The average speed of the photoelectron is around $2.4$ nm/fs, while the interaction time can be considered equal to the XUV pulse duration. 
An approximate classical limit is given by $2.4 ~\mathrm{nm/fs} \times 73~\mathrm{fs}=175$ nm.
The wavepacket describing the entangled photoelectron can therefore expand to a mesoscopic scale, while the ion remains coupled to the field.
Evidently, our experiment demonstrates non-locality in photoionization experiments, wherein two massive quantum objects remain entangled despite the wavepacket corresponding to one of them getting spread over hundreds of nanometer. In comparison, macroscopic entanglement between two trapped ion-pairs have been shown to exist across a distance of $240~\mu$m\cite{jost2009}. There, the ion-pairs were slowly moved apart and then back together again, while being cooled down to micro-Kelvins. In our experiment, the emitted electron wavepacket is expanding freely in space at room temperature, and it is not brought back together or, recombined with its parent ion to reveal the entanglement.
The experimental spectra in Fig.~2(A and B) are accumulated over more than $10^5$ laser-shots. The loss of entanglement for some of the laser-shots can mostly be attributed to the fluctuations of the temporal coherence of the FEL pulse. Several factors such as, imperfections in the seeded beam, or in the electron bunch trajectory, and chirp of the XUV pulse, can reduce the temporal coherence of the outgoing XUV beam. This is visible in the observed shot-to-shot variation of the photon bandwidth (see SM for details), which, for a fixed value of the pulse duration, would imply wide deviations from the the Fourier-transform limit.

For $I^\mathrm{FEL}=1.25\times10^{13}$ W/cm$^2$ and $\Delta t^\mathrm{FEL}=73$ fs, we are in the non-perturbative coupled ion-field regime (see Fig.~1B). Here, the total ion population grows linearly with the interaction time, as predicted by Fermi's Golden rule, but the population is shared in both ion channels with an oscillatory behaviour (see Fig.~4A). These modulations, predicted previously in refs.\cite{grobe1993,zhang2014}, evolve at the same rate as the Rabi oscillations, with period $T_R$. Our analytical model shows that they result from a convolution of the atomic photoionization effect with subsequent ionic Rabi oscillations. The duration of the pulse in our experiment is close to $\Delta t^\mathrm{FEL}\approx 3.4T_R$, which results in six clear crossings of such modulations of the ionic populations. At resonance, with $\omega=\omega_{ba}$, the photoelectrons exhibits symmetric doublets in both ionic channels (see Fig.~4B). We label the photoelectron in the higher (lower) energy peak as $E^\mathrm{kin}_+$ ($E^\mathrm{kin}_-$). The fact that each ion state is coupled to a superposition of two non-degenerate photoelectron continuum peaks implies that the energy conservation between stationary states is not sufficient and that a dynamical picture is required. We employ the dressed-state picture, where uncoupled ion states are associated to Fock states for the field as: $|a,N-1\rangle$ and $|b,N-2\rangle$, with $N$ being the initial number of photons associated with a neutral atom: $|g,N\rangle$ (see SM for the quantum optics formalism). The model is in excellent agreement with the numerical simulations based on the time-dependent Dirac equation (see Fig.~4B). 
Further insights into the ion-photoelectron entanglement can be found from the relative phase between the two ionic states coupled to an identical photoelectron state. In Fig.~4C, the relative phases are separately shown for each photoelectron peak. The photoelectron peak with $E_-^\mathrm{kin}$ ($E_+^\mathrm{kin}$) has a phase difference that approaches $\pi/2$ ($-\pi/2$). Thus, the final state approaches:  
\begin{equation}
|\Psi^{({\cal{F}})}\rangle \sim \frac{1}{2}\left[(|a,N-1\rangle+i|b,N-2\rangle)\otimes|E^\mathrm{kin}_-\rangle+(|a,N-1\rangle-i|b,N-2\rangle)\otimes|E^\mathrm{kin}_+\rangle \right],
\end{equation}
which is fully entangled because the superposition of the ion states are orthogonal due to dressing by the field, and the two continuum states are also orthogonal due to their different kinetic energies, $E^\mathrm{kin}_-\ne E^\mathrm{kin}_+$. Small phase effects due to AC Stark shifts and spin-orbit interaction are identified in Fig.~4C in the numerical results based on the Dirac equation.

Finally, the degree of entanglement of the dressed ion and the photoelectron can be theoretically examined, for the case of a single atom (pure state) interacting with a laser field, using the von Neumann entropy of entanglement, $S=-\mathrm{Tr}[\rho_{\cal P}\log_2(\rho_{\cal P})]$. This depends on the reduced density matrix of the photoelectron, $\rho_{\cal P}$, conditioned on the photoionization event\cite{haroche2006,ruberti2022}. Despite being a difficult quantity to estimate\cite{islam2015,brydges2020}, the von Neumann entropy is a quantitative measure of the lack of knowledge of the ionic (electronic) system due to its entanglement with an unresolved photoelectron (ion): $S=S_{\cal P}=S_{\cal I}$. Experimentally, photoelectrons from He atoms were collected without measuring in coincidence the state of the He$^+$ ion. This corresponds to measuring the diagonal elements of the reduced density matrix of the photoelectron: $\rho_{\cal P}(E^\mathrm{kin},E^\mathrm{kin})=|\alpha_{a}(E^\mathrm{kin})|^2+|\alpha_{b}(E^\mathrm{kin})|^2$. The analytical model is used to construct the time-dependent density matrix of the process. The entropy of entanglement is shown in Fig.~4D as a function of the interaction time. It is observed that the entropy rises from zero to its maximally allowed non-relativistic (qubit) value of $\log_2(2)=1$ within the interaction time of the experiment in the resonant case. The entropy is found to be modulated at larger interaction times, but it tends to its maximal value with reduced modulations over time. The entanglement entropy is close to $1$ beyond an interaction time of $20$~fs, implying that the photoelectron wavepacket remains fully entangled to the dressed ion during the entire time from $20$ fs to $73$ fs, despite being delocalized across mesoscopic length-scales. We interpret this result as a requirement of one Rabi cycle for entanglement to manifest itself between electron and ion.
In other words, the generation of entanglement between the photoelectron and the dressed ion does not take place immediately, rather it can be delayed by a few femtoseconds following emission of the photoelectron. Relativistically, the behaviour of the entropy is very similar and it is not found to exceed the non-relativistic value despite the larger populated ion subspace, constituting a qutrit ($1s_{1/2}\leftrightarrow \{2p_{1/2},2p_{3/2}\}$), allowing a maximal value of $\log_2(3)\approx 1.58$ entropy of entanglement. The excellent agreement between the analytical model and the numerical simulations shows that many-level strong-field effects beyond our model are not essential under the conditions of our experiment; see Fig.~4(A-D).

Our results show that with femtosecond XUV pulses from a seeded FEL one can create a unique bipartite system: the photoelectron and the dressed ion. The use of a short wavelength (thus, high energy) photon pulse as an entangler allowed us to generate entanglement between two massive particles across mesoscopic length-scale at room temperature. Despite the electron wavepacket spreading across distances of more than $100$ nm within the interaction time of few tens of femtoseconds, the entanglement persists, providing direct proof of non-locality in photoionization processes across ultrafast timescales. Using the time-dependent von Neumann entropy, we have illustrated the evolution of quantum entanglement for hybrid states of matter. We anticipate that similar approaches can be useful in studying the dynamical evolution of entanglement in multi-electronic systems\cite{ishikawa2023}. In the future one can employ quantum state tomography protocols to access similar information experimentally\cite{laurell2022}, thereby allowing us to quantify entanglement in photoionization. By reducing the wavelength of the entangler, one can access electrons from atomic core levels, opening up the possibility to study entanglement between Auger- and photoelectrons\cite{bostrom2018}. Free electrons with higher kinetic energies could provide opportunities to investigate entanglement between wavepackets delocalized across macroscopic length-scales. Our experimental scheme is general enough so that it can be extended to multi-center systems, such as molecules, following site-selective ionization by intense short-wavelength pulses from FELs. As seeded free-electron lasers providing highly coherent X-rays are becoming progressively available\cite{nam2021}, we believe the present study can encourage future experiments concerning entanglement between massive particles physically separated by large distances.

\section*{Materials and Methods}
\paragraph*{Experimental Methods}

The experiment was performed at the Low Density Matter (LDM) beamline\cite{svetina2015}, which contains a Kirkpatrick-Baez system of two mirrors to focus the linearly polarized XUV light from the FEL undulator down to $6~\mu$m, allowing one to reach instaneous intensities above $10^{14}$ W/cm$^2$ at the best focus. We used a pulsed Even-Lavie valve, operating in synchronization with the arrival of the FEL pulses (repetition rate: $50$ Hz), to deliver the target atoms in the form of a supersonic jet. The number density of the helium atoms in the interaction region was estimated to be around $10^{22}$/m$^3$. It leads to a mean-free-path of about $1.1$ mm between two helium atoms, substantially larger compared to the de-Broglie wavelength of the emitted photoelectron ($\sim 0.3$ nm), thus preserving the single collision condition. A $2$-m long magnetic bottle electron spectrometer (MBES) was mounted perpendicular to the plane of the interaction. The photoelectrons were retarded electrostatically down to $1$ eV kinetic energy enabling the resolution of the MBES to be $E/\Delta E \approx 50$. We calibrated the kinetic energy scale by the spin-orbit splitting in valence-ionized Argon atoms. The wavelength of the FEL radiation was varied from $30.2$ to $30.4$ nm and back, during the measurement of the photoelectron spectra. In order to collect intensity dependent photoelectron spectra, the measurements were repeated with aluminum filters (thickness: $50$ nm and $770$ nm) in the beam. The FEL undulator was set at the $8^{\mathrm{th}}$ harmonic of the seed laser operating at $243$ nm. We found the variation of the experimental bandwidth of the XUV pulse to be approximately between $20$ and $76$ meV (see SM for details). The value of the pulse duration (FWHM), and the photon bandwidth, as obtained from simulations using PERSEO\cite{giannessi2006}, were on average $73$ fs and $37$ meV, respectively (see SM for details). Similarly, the PERSEO-simulated value of the group-delay-dispersion for the FEL pulses was found to be $+325$ fs$^2$. Immediately after the undulator, the energy per pulse for the FEL beam was measured to be around $57~\mu$J. At the best focus, the beam diameter  was assumed to be $4\sigma$, where $\sigma=6/2.355\approx2.55~\mu$m. This led to a beam waist, $w_0=2\sigma=5.1~\mu$m and subsequently, a Rayleigh length of $2.7$ mm. 

\paragraph*{Theoretical Methods}

In order to understand the non-linear dynamics of the He atom in the intense FEL field, we use a quantum optics framework based on the resolvent-operator technique\cite{tannoudji2004}. Assuming that the interaction starts at $t_0=0$ with the atom in its ground state, $\left|g,N\right>$, the total time-dependent wavefunction of the He atom and FEL field at time $t>0$ becomes (see SM for details):   
\begin{displaymath} 
\left|\Psi (t)\right>  
    =  g(t) e^{-\mathrm i\tilde \epsilon_g t} \left| g,N \right>  
    + \frac{\Omega^{ag}}{2}\int_{0}^{t} dt' \int d E^\mathrm{kin} g(t') e^{-\mathrm i \tilde \epsilon_g t'}    
\end{displaymath}
\begin{equation}
     \times \left( 
    a(t-t') e^{-\mathrm i \tilde\epsilon_a (t-t')} \left| a,E^\mathrm{kin},N-1 \right> + 
    b(t-t') e^{-\mathrm i \tilde\epsilon_b (t-t')} \left| b,E^\mathrm{kin},N-2 \right> 
    \right), 
\end{equation}
where $\tilde\epsilon_i$ are energies of the associated uncoupled atom--field states. The first term describes depletion of the He atom by the FEL field at the rate $1/\tau_g$, i.e. $g(t)=e^{-t/2\tau_g}$, and the second term describes the transition, at time $t'$, to the He$^+$ ion and photoelectron bipartite system. Within the rotating-wave approximation, the subsequent ionic dynamics are given by well-known Rabi amplitudes: $a(t-t')$ and $b(t-t')$, with $a(0)=1$, $b(0)=0$ and $t\ge t'$. The photoelectron is assumed to be freely propagating after the transition time, $t \ge t'$. Integration over $t'$ gives us analytical expressions for the complex amplitudes of the uncoupled photoelectron--ion--field states: $c_a(t,E^\mathrm{kin})$ and  $c_b(t,E^\mathrm{kin})$. We then construct the time-dependent density matrix, conditioned on the photoionization event, which allows us to compute the von Neumann entropy of entanglement between ion and photoelectron. Spin is treated within the non-relativistic (singlet) approximation in our model. In order to verify our non-relativistic model, we have numerically computed the interaction amplitudes, $g(t)$, $a(t)$ and $b(t)$ for the He atom and He$^{+}$ ion using the time-dependent Dirac equation within the semi-classical approximation, including complete sets of electronic states, where the excited ionic state $2p^+$ is split up into two relativistic states: $2p^+_{j=1/2}$ and $2p^+_{j=3/2}$. Our results show that strong-field effects and relativistic effects are of the same magnitude under the experimental conditions, but that does not alter the formation of ion-electron entanglement.         

\bibliographystyle{Science} 
\bibliography{myref}

\paragraph*{Acknowledgments}
We acknowledge the assistance of the staff at FERMI during the beamtime. 

\paragraph*{Funding} We were supported by the project Laserlab-Europe under Grant Agreement 654148 from the EU Seventh Framework Programme (FP7/2007-2013). S.N. thanks Centre National de la Recherche Scientifique (CNRS) and F\'{e}d\'{e}ration de Recherche Andr\'{e} Marie Amp\`{e}re, Lyon for financial support. J.M.D. is funded by the Knut and Alice Wallenberg Foundation (2017.0104 and 2019.0154), the Swedish Research Council (2018-03845) and the Olle Engkvist's Foundation (194-0734). R.F. acknowledges financial support from the Swedish Research Council (2018-03731) and the Knut and Alice Wallenberg Foundation (2017.0104). P.E.-J. acknowledges support from the Swedish Research Council (2017-04106) and the Swedish Foundation for Strategic Research (FFL12-0101). 

\paragraph{Author Contributions} D.B., C.C., M. Di F., P.E.-J., R.F., G.G., M.G., J.P., O.P., K.C.P., R.J.S., S.Z. and S.N. carried out the experiment, collected the data and contributed to the preliminary analysis of the data. The theoretical investigation was conducted by A.S., A.P., E.O., M.B., S.C., and F.Z., under the supervision of J.M.D. The analytical model was developed by A.S. while A.P. carried out the relativistic calculations. S.N. performed the in-depth data analysis. M. Di F., O.P. and C.C. were responsible for the LDM end-station. R.J.S. and R.F. provided the magnetic bottle electron spectrometer. L.B., M.B.D., and L.G. optimized the accelerator and provided the FEL beam. M.Ma., M.Z. characterized the FEL pulses. P.V.D. contributed to the initial analysis of the experimental data. M.Me. and C.M. helped with the planning of the experiment. J.M.D. and S.N. wrote the manuscript, which all authors discussed. S.N. proposed and led the project.

\paragraph{Competing Interests} The authors declare that they have no competing interests.

\paragraph{Data and Materials Availability} All data needed to evaluate the conclusions in the paper are present in the paper and/or the Supplementary Materials. 

\clearpage

\begin{figure*}[!htb]
\centering
\includegraphics[width=1.0\linewidth]{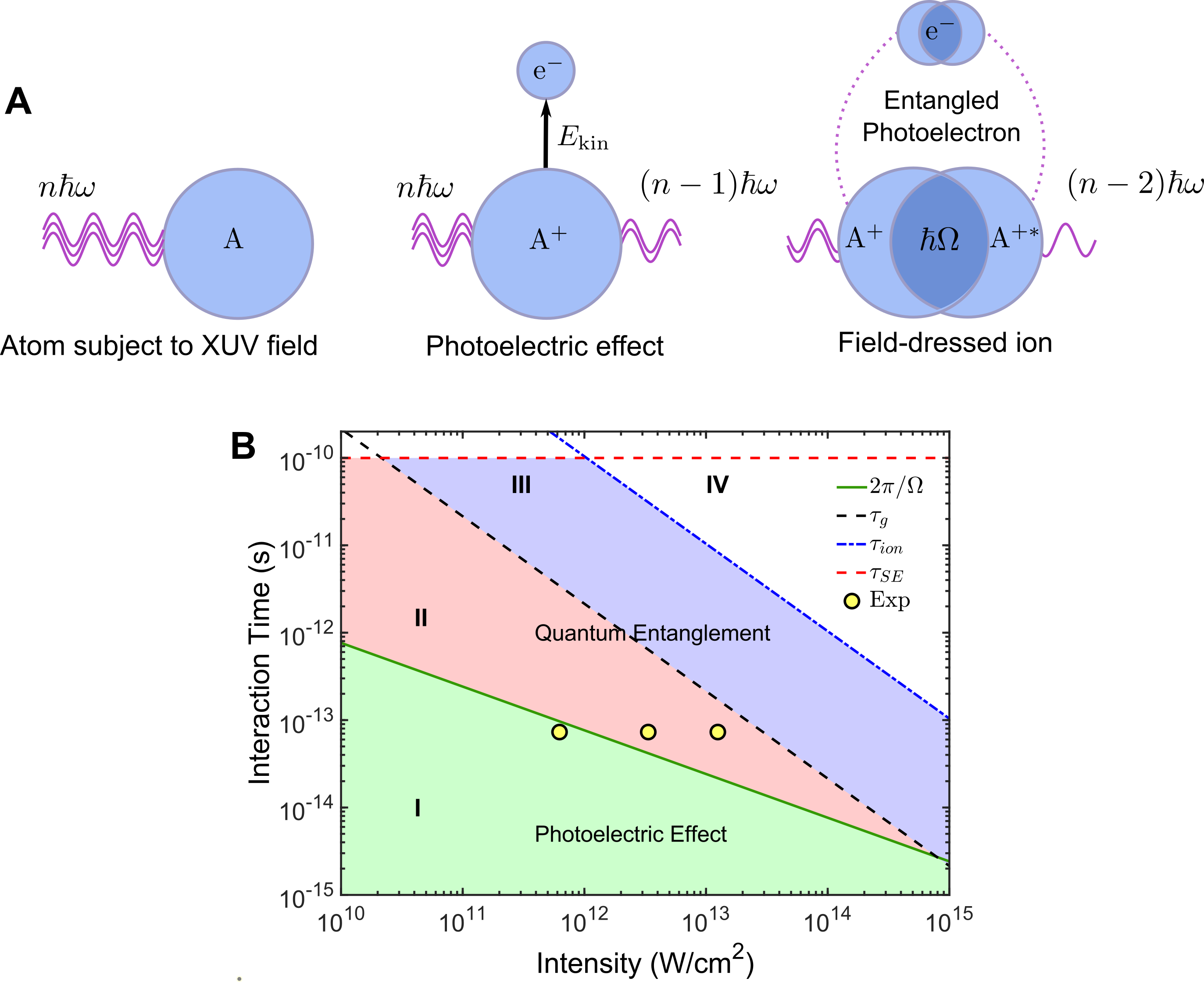}
\caption{{\bf Quantum entanglement between photoelectron and dressed ion.} 
({\bf A}) Atoms (denoted A) subjected to light may eject electrons (e$^-$) with a particular kinetic energy ($E_\mathrm{kin}$) that depends on the frequency of the light ($\omega$). However, if the photon energy ($\hbar\omega$) matches a transition in the ion (A$^+\rightarrow$A$^{+*}$), a periodic energy exchange may follow at a rate given by the Rabi frequency ($\Omega$).  Due to the formation of such ``dressed states'' of light and matter, quantum entanglement between the ion and the emitted photoelectron can be studied using intense high-frequency FEL pulses. ({\bf B}) In the Region I, the interaction time is shorter than the Rabi period ($<2\pi/\Omega$). In contrast, in the Region II, the interaction time is sufficiently long to dress the ion and entangle the photoelectron ($>2\pi/\Omega$). In Region III, the neutral atomic population is saturated ($>\tau_g$), while in Region IV the dressed ion population is lost to further photoionization ($>\tau_{\mathrm{ion}}$). Coherent processes in the ion are limited by the spontaneous emission lifetime ($\tau_{\mathrm{SE}}$). The experiments were performed at three different intensities, allowing us to study the generation of quantum entanglement via the photoelectron spectra.}
\label{fig1}
\end{figure*}

\begin{figure*}[!htb]
\centering
\includegraphics[width=0.94\linewidth]{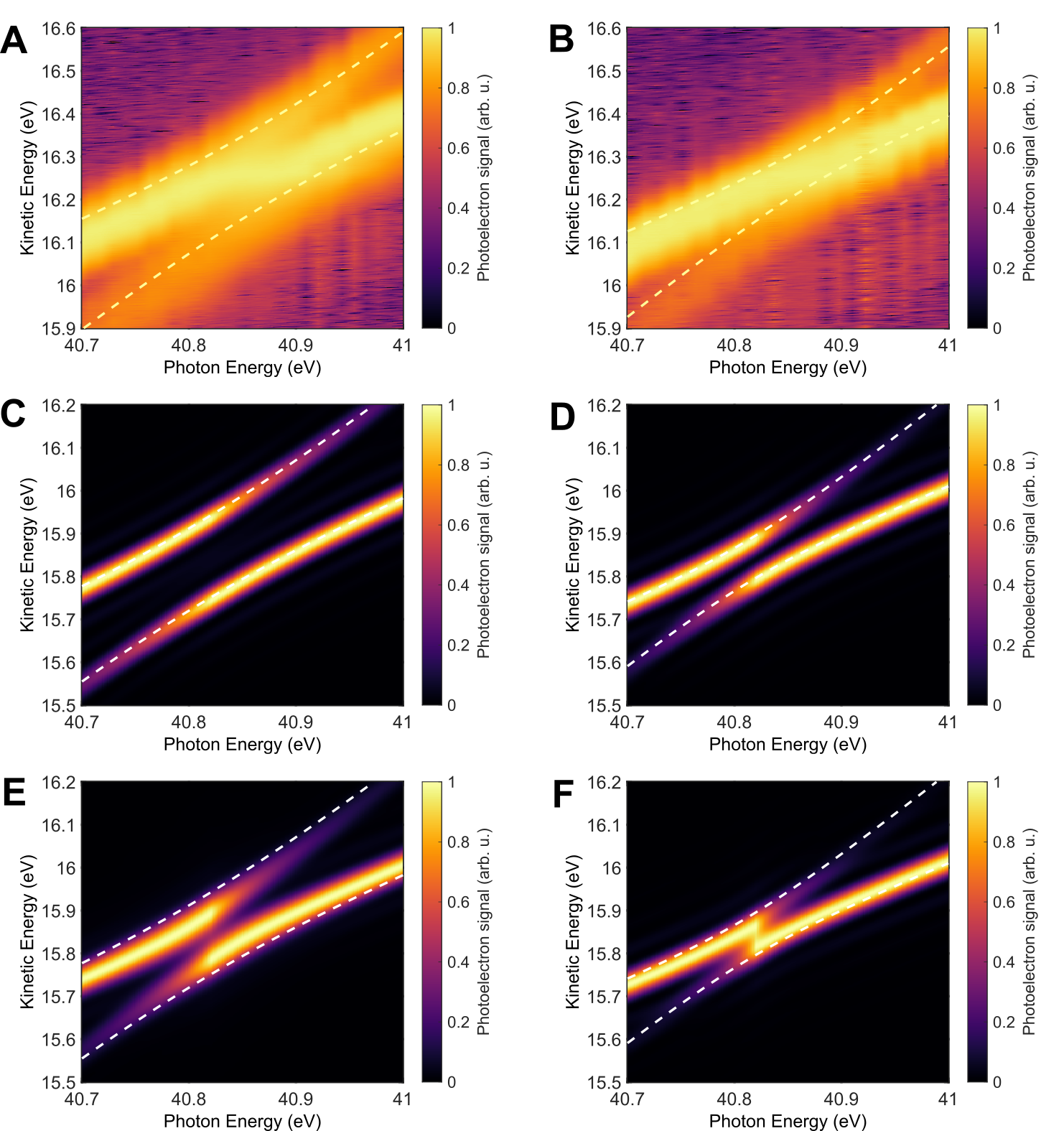}
\caption{{\bf Avoided crossing in the photoelectron spectra.} ({\bf A}) Measured photoelectron spectra (shown in logarithmic scale) as a function of photon energy across the $1s\rightarrow 2p$ transition in He$^+$ ion at an effective intensity of $1.25 \times 10^{13}$ W/cm$^2$. ({\bf B}) Same as ({\bf A}), but at an intermediate effective intensity of $3.35 \times 10^{12}$ W/cm$^2$ for the XUV-FEL pulse. ({\bf C}) Photoelectron spectra from the analytical model (shown in linear scale) for an XUV-FEL pulse with the same intensity as that in ({\bf A}). ({\bf D}) Same as in ({\bf C}) but with an intensity equal to that in ({\bf B}). ({\bf E, F}) Macroscopically averaged photoelectron spectra for two different intensities. In each panel, the dashed lines correspond to dressed-state energies with the binding energies being calibrated with respect to ref.\cite{kramida2022} (for experiment) or to ref.\cite{zapata2022} (for theory).}
\label{fig2}
\end{figure*}

\begin{figure*}[!htb]
\centering
\includegraphics[width=1.0\linewidth]{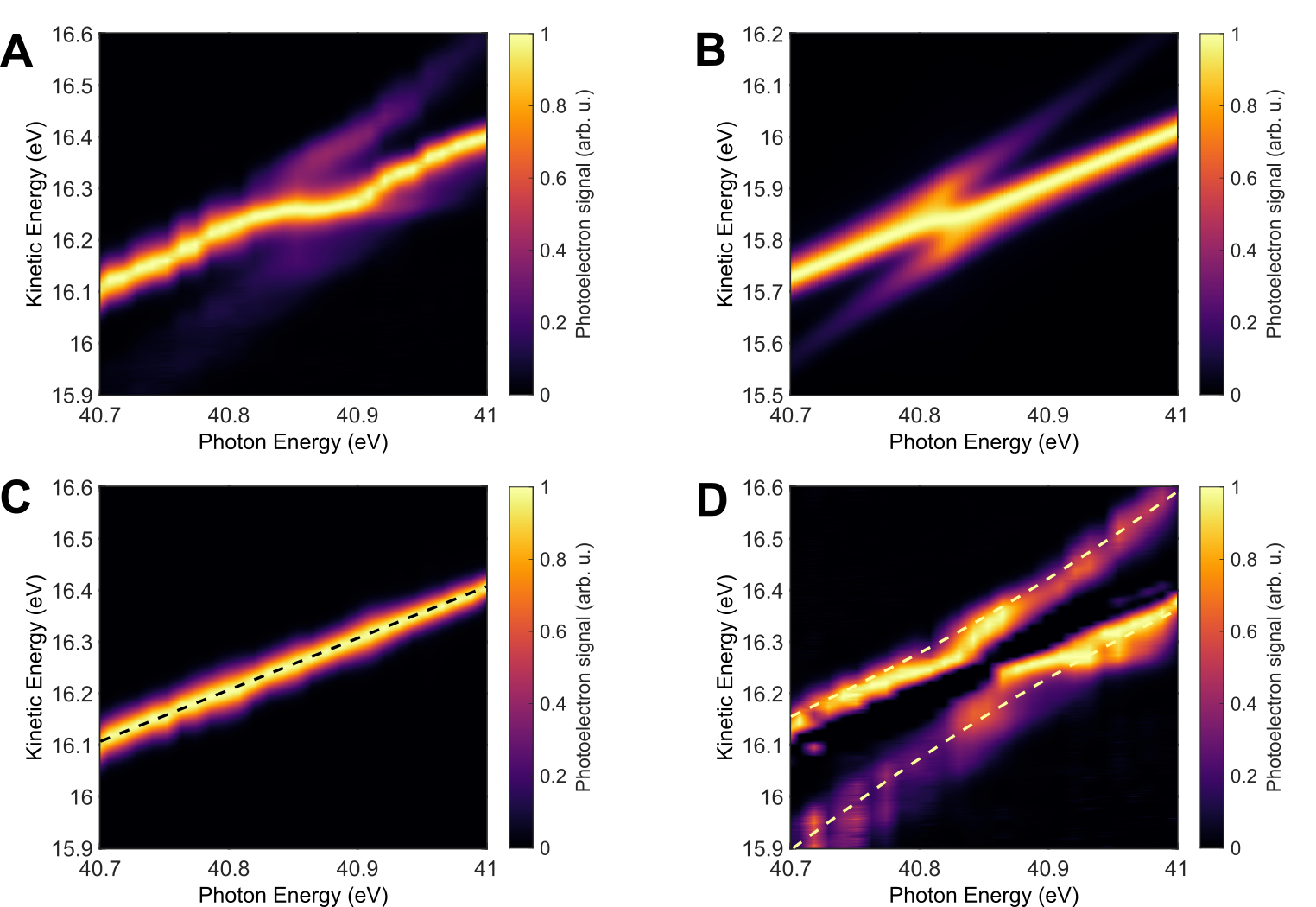}
\caption{{\bf Entangled and non-entangled photoelectron spectra.} ({\bf A}) Deconvoluted experimental photoelectron spectra at an effective intensity of $1.25 \times 10^{13}$ W/cm$^2$ for the XUV-FEL pulse. ({\bf B}) Macroscopically averaged photoelectron spectra at the same intensity from the analytical model, taking into account both entangled and non-entangled contributions in near equal amounts. ({\bf C}) Experimental photoelectron spectra without any entanglement. The dashed line corresponds to the difference between the photon energy and the binding energy of neutral helium. ({\bf D}) Photoelectron spectra with quantum entanglement between the electron and the dressed-state of the ion, obtained by taking a difference between ({\bf A}) and ({\bf C}). The dashed lines once again denote the corresponding dressed-state energies; same as in Fig. \ref{fig2}A.}
\label{fig3}
\end{figure*}

\begin{figure*}[!htb]
\centering
\includegraphics[width=1.0\linewidth]{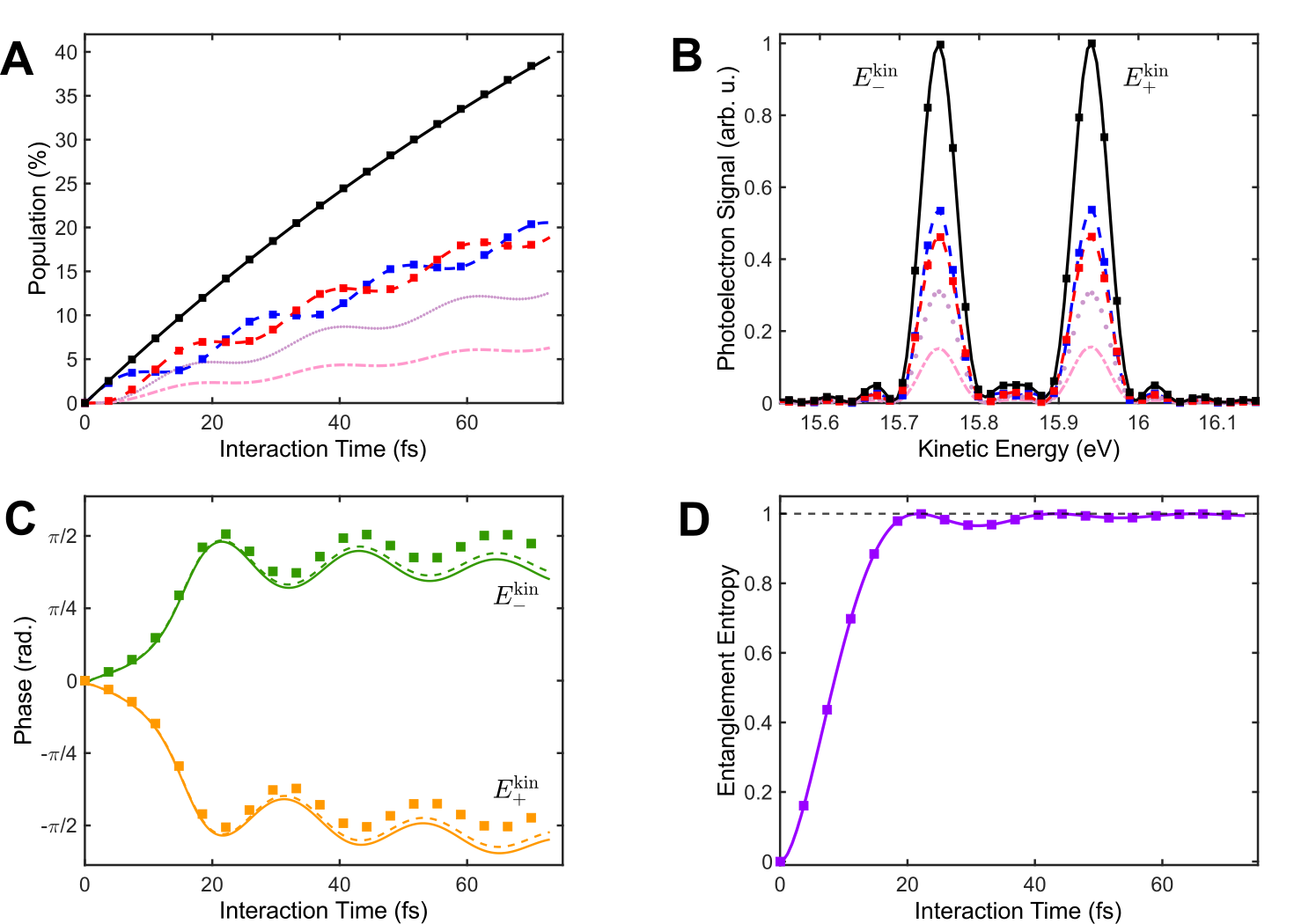}
\caption{{\bf Theoretical interpretation.} ({\bf A}) The time-dependent ionic populations calculated by the Dirac equation (lines) and analytical model (solid squares): $1s$ in He$^+$ (blue dashed line and blue squares), $2p$ in He$^+$ (red dashed line and red squares), and total population in both $1s$ and $2p$ (black solid line and black squares). Individual contributions from two spin-orbit split levels, $2p_{1/2}$ (dashed-dotted line) and $2p_{3/2}$ (dotted line) are also shown. ({\bf B}) Same as ({\bf A}), but for the photoelectron spectra at the end of the pulse for different ionic levels. ({\bf C}) Difference in phase between the contributions of $1s$ and $2p_{1/2}$ levels (green solid line) in He$^+$, as calculated by the Dirac equation, for the peak at lower kinetic energy, $E_-^{\mathrm{kin}}$. The same between $1s$ and $2p_{3/2}$ is also shown (green dashed line). For the peak at higher (lower) kinetic energy, $E_+^{\mathrm{kin}}$ ($E_-^{\mathrm{kin}}$), the corresponding phase differences are negative (positive); the solid (dashed) yellow line indicates the difference between $1s$ and $2p_{1/2}$ $\left(2p_{3/2}\right)$. In both cases, the solid squares denote the phase differences $1s$ and $2p$ levels in He$^+$ as obtained by the analytical model. ({\bf D}) von Neumann entropy of entanglement as a function of interaction time, by Dirac equation (solid line) and the analytical model (solid squares).}
\label{fig4}
\end{figure*}

\end{document}